\documentstyle[preprint,epsf,prl,aps]{revtex}

\begin{document}
\draft
\title{
The Kosterlitz-Thouless transition in the XY Kagom\'{e} antiferromagnet}
\author{ 
V.~B.~Cherepanov $^{a,b,c}$,
I.~V.~Kolokolov $^{a,d}$, and
E.~V.~Podivilov $^{a,c}$, 
}
\address{
$^a$ Department of Physics of Complex Systems, Weizmann
Institute of Science, Rehovot, 76100, Israel \\
$^b$ School of Physics and Astronomy,  The Sackler
Faculty of Exact Sciences, Tel~Aviv University, Tel~Aviv, 69978,
Israel \\
$^c$ Institute of Automation and
Electrometry, 630090, Novosibirsk, Russia \\
$^d$ Budker Institute for Nuclear Physics, 630090,
Novosibirsk, Russia}

\date{\today}
\maketitle
\widetext
\begin{abstract}
The problem of  the Kosterlitz-Thouless (KT) transition in the highly
frustrated XY 
Kagom\'{e}  antiferromagnet  is solved. The problem is mapped onto that of
the KT transition in the XY ferromagnet on the hexagonal lattice. The
transition temperature is found. It is shown that the spin correlation
function  exponentially decays with distance even in the
low-temperature phase, in contrast to the order parameter correlation
function, which decays algebraically with distance.

\end{abstract}


\narrowtext

Generally, XY spins on two-dimensional lattices undergo a
Kosterlitz-Thouless (KT) transition \cite{B,KT}. As a rule, physics of
this transition does not depend on details of lattice structure. In
the low temperature phase, pairs of spin direction singularities,
vortices, with opposite topological  charges form
quasi-molecules. Above the transition temperature, the quasi-molecules
decay into a vortex gas.  In the low-temperature phase, the spin
correlations decay algebraically with 
distance, above the transition temperature, they decay exponentially.
Formally, the universality of the KT transition follows from the
possibility to describe low-energy states of two-dimensional XY spin
systems in terms of the nonlinear $\sigma $-model.
The XY antiferromagnet on the two-dimensional Kagom\'{e} lattice is an 
exception from this class. It has infinitely many ground states,
therefore its low-temperature properties cannot be described by the
nonlinear $\sigma $-model. This makes the problem of a KT-like
transition in the XY Kagom\'{e} antiferromagnet a special one, which
is substantially more complicated than the standard theory of the KT
transition. The very possibility of a KT transition in such an unusual
system is under question. 

The problem of the KT transition in the Kagom\'{e} antiferromagnet was first
addressed by Huse and Rutenberg \cite{HR}. They suggested that the
order parameter for the KT transition is $ e^{ 3 i \theta} $ where
$\theta $ is the angle of a spin. This order parameter is invariant
with respect to any arbitrary choice of ground states, which are a
subset of local $ 2 \pi/3 $ spin rotations. Therefore this order
parameter can change smoothly in the plane even though spins rotate
locally at multiple $ 2 \pi/3 $ angles. An indirect evidence of the KT
transition in the Kagom\'{e} antiferromagnet was obtained from MC
simulation \cite{R}.

A network of Josephson junctions with the $\pi $-phase
shift can be mapped onto the antiferromagnetic XY model as 
well. Experimental studies of artificial networks of Josephson 
junctions on the Kagom\'{e} lattice  make this problem especially
appealing \cite{DM}.

In this paper, we examine the KT-like
transition in the XY Kagom\'{e} antiferromagnet. We find that the KT
transition in the Kagom\'{e} antiferromagnet does exist, we evaluate
the transition temperature, and we show that the spin correlation
function behaves itself in an unusual way. 
\begin{figure}

\epsfxsize = \hsize
\epsffile{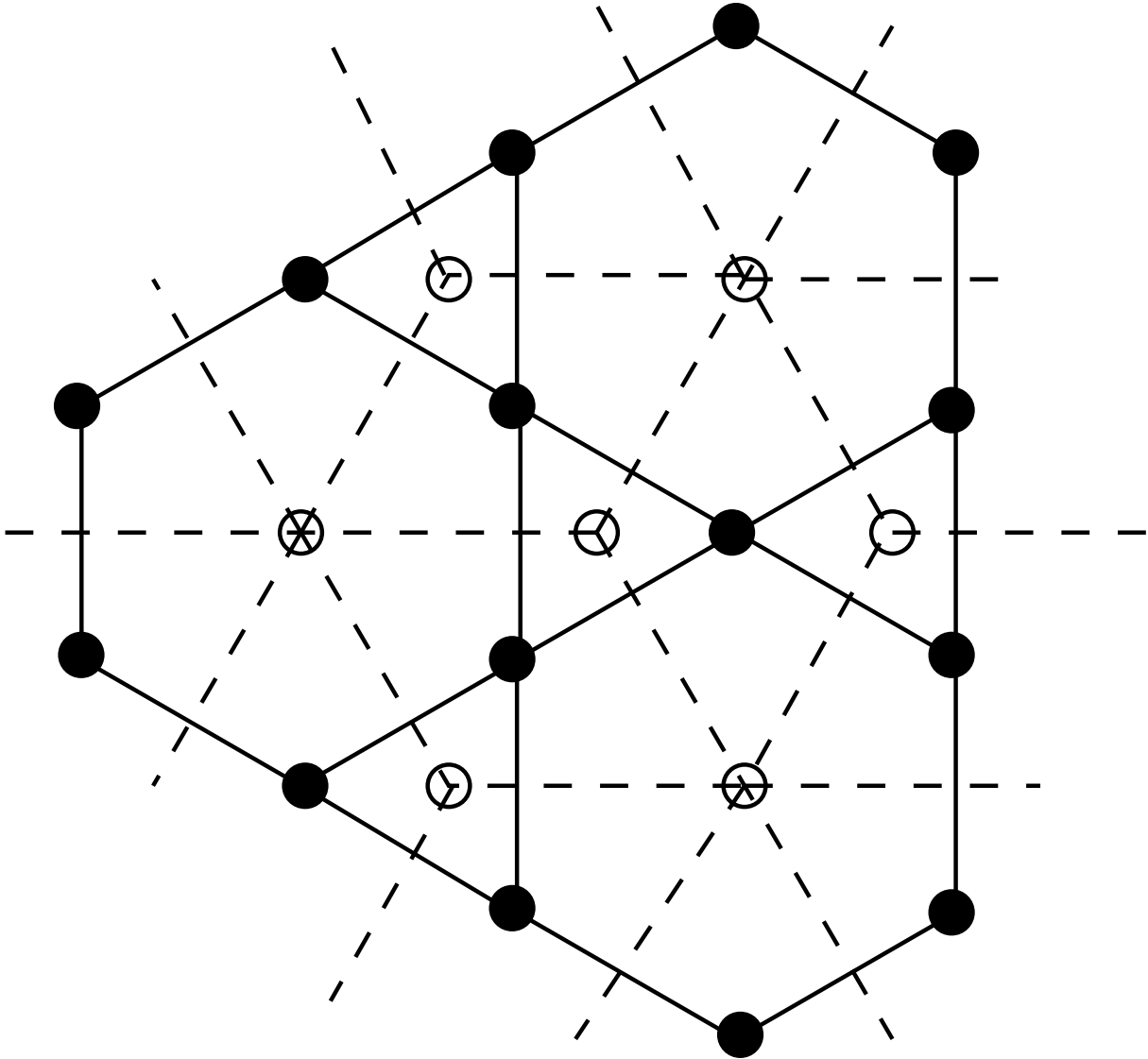}

\caption{The Kagom\'{e} lattice (filled dots) with antiferromagnetic bonds
(continuous lines)
and the dual lattice
(circles) and its bonds (dashed lines). }

\end{figure}

The common  treatment of the KT transition in the continuous limit
\cite{CL,N} does not account for the  
structure and high degeneracy of the system and therefore it is not
applicable in this case.
In order to take into account the special structure of the Kagom\'{e}
lattice we follow the lattice approach developed by Jose, Kadanoff,
Kirkpatrick, and Nelson (JKKN) \cite{JKKN}. The Kagom\'{e} lattice
consists of triangles and hexagons (fig.~1). 
The Hamiltonian   of the Kagom\'{e} antiferromagnet  can be
represented as a sum of squares of the total spins in triangles of the
nearest neighbors, ${\bf S}_{\Delta}$: 
\begin{equation}
H = 1/2 \; J \sum_{\Delta} \left( {\bf S}_{\Delta}
\right)^{2},
\label{a0}
\end{equation}
where $\Delta$ numbers the triangles of the nearest neighbors
(fig.~1).  Each spin participates in two triangles. 
The ground state energy is equal to zero and there
are  infinitely many ground states with ${\bf S}_{\Delta} = 0 $. 
In any ground state, the angles between neighboring spins are equal to
$\pm {2 \pi}/3$. Vortices are structural defects in the ground
state spin pattern with kernels localized in the centers of hexagons
or triangles.  

Our plan is as follows.
Following JKKN, we represent the
partition function in terms of integer-valued currents on the
lattice and introduce the dual lattice to the Kagom\'{e} lattice, 
with sites located in the centers of triangles and hexagons. 
A very important step is to exclude vortices located in the centers of 
triangles 
from further consideration because the energy of such vortices is
approximately $J \ln {(2 \sqrt{3} )}$ higher than that of vortices
residing on 
hexagons. The neglect of such vortices 
makes sense because, as we show further (Eq.~(\ref{r})), 
the KT transition temperature, $T_{c} = 0.0756 J$
is small compared to  the energy of a vortex residing on a triangle,
and the contribution of such vortices is of order 
$  {(2 \sqrt{3} )}^{- J/{T_{c}}} \approx 7\cdot 10^{-8} \ll 1$.
This small parameter allows one to integrate out currents in 
triangles in the partition function. After this step, only 
currents in hexagons and new variables, chiralities,
located on triangles and equal to $\pm 1$ are left. The chirality
distinguishes 
the clockwise and counter-clockwise configurations of spins in each
triangle. 
We show that summation over 
chiralities renormalizes the partition function expressed in terms of
currents in hexagons, however 
it does not change the low-temperature partition function
drastically. We find that the degeneracy of ground states results in a
decrease of the transition temperature compared to that of the
XY ferromagnet on the hexagonal lattice. The order parameter
correlations below the KT transition temperature decay as a power of
distance. We show that in the Kagom\'{e} antiferromagnet the
spin-spin correlations decay exponentially, in contrast with
those in ``usual'', non-degenerate XY magnets. The reason for 
such a fast decay of the spin-spin correlations is that the symmetry of
the pair spin correlation function is lower than that of the order
parameter and those correlations are sensitive to the multiplicity of
ground states, that are mainly disordered.

The partition function of the XY Kagom\'{e} antiferromagnet can be
represented as a sum over the lattice's bonds
\begin{equation}
Z({\beta}) = \int 
e^{- \beta \sum_{\bf r,a}
\cos{ \left[ \theta{({\bf r})} - \theta{({\bf {r + a}})}
\right] } }
\prod_{\bf r} {\rm d} \theta{({\bf r})},
\label{a}
\end{equation}
where ${\bf r}$ marks positions on the Kagom\'{e} lattice, ${\bf a}$
are the three lattice vectors directed along the antiferromagnetic
bonds between nearest neighbors, $\theta_{\bf r}$ are the spin angles,
and $\beta = J S^2/2T$ is
the  dimensionless inverse temperature.
The $2 \pi$ periodicity of the angle variables allows one to expand
$ Z({\beta}) $ in Fourier series with the coefficients 
$I_{n{({\bf r,a})}}(- \beta )$, where  $ I_{n}(x)$ is the modified
Bessel function and  integer numbers $ n({\bf r,a})$ are located on
bonds connecting nearest neighbors ${\bf r}$ and ${\bf {r + a}}$. 

One can integrate over the angles 
$\theta{({\bf r})}$ in the partition function.
This results in the following
representation for the partition function:
\begin{equation}
Z( \beta ) = \sum_{\{n({\bf r,a}) \}} \prod_{(\bf {r, a} )} 
I_{n({\bf r,a})}( - \beta ) 
 \delta
\left( \sum_{\bf {a = r - r'}} n({\bf r,a}) \right),
  \label{d}
\end{equation}
where 
$  n({\bf r - a, - a}) = -  n({\bf r,a}).$
The numbers $ n({\bf r,a})$ obey the conservation condition  at each
site of the lattice:
\begin{equation}
\sum_{\bf a} n({\bf r,a}) = 0.
  \label{f}
\end{equation}
Following JKKN \cite{JKKN}, we introduce the dual lattice to the
Kagom\'{e} lattice in order to account for the conservation conditions
(\ref{f}). The sites of the dual lattice, ${\bf R}$, are located at
crossings of lines perpendicular to the bonds that connect nearest
neighbors on the Kagom\'{e} lattice and cross the bonds in their
middle points  (fig.~1). This construction provides a one-to-one
correspondence between bonds of the Kagom\'{e} lattice and bonds of 
the dual lattice.
The conservation conditions (\ref{f}) can be resolved if one
introduces
integer-valued currents $J({\bf R})$ circulating in each triangle and
in each hexagon of the the Kagom\'{e} lattice. The currents can be
assigned to sites ${\bf R}$ of the dual lattice located  in
the centers of triangles and hexagons. A current along a bond ${\bf
  (r,a)}$, $n({\bf r,a})$, is equal to the sum of currents in one
triangle and in one hexagon that share the bond ${\bf (r,a)}$. The
currents $J({\bf R})$ obey the Kirchhoff's rule: the sum of currents
arriving at each site is equal to the sum of departing ones. Thus, the
conservation conditions (\ref{f}) are fulfilled through the
Kirchhoff's rules.

The summation over currents $ \{ n({\bf r,a}) \} $ with the
conservation laws (\ref{f}) in the partition function (\ref{d}) is
equivalent to summation over currents $ \{ J({\bf R}) \}$ located on
sites of the dual lattice
\begin{equation}
Z({\beta}) = \sum_{\{ J({\bf R}) \} } \prod_{\bf (R,A)}
I_{J({\bf R + A}) + J({\bf R})}( - \beta ),
  \label{g}
\end{equation}
where ${\bf A}$ are the lattice vectors of the dual lattice. 

From this point, our way deviates from that by JKKN. The
product $ \prod_{\bf (R,A)} $ over the bonds can be factorized as a
product over all triangles of the  Kagom\'{e} lattice and triple
products of the Bessel's functions corresponding to the bonds in each
triangle because each bond of the  Kagom\'{e}
lattice belongs to one and only one triangle. This allows one to
represent the partition function as follows:
\begin{equation}
Z({\beta}) = \sum_{\{ J({\bf R_{h}}) \} } \prod_{\bf (R_{t})}
\sum_{ J({\bf R_{t}})} \prod_{h' = 1}^{3}
 I_{J({\bf R_{t} + A_{h'} }) + J({\bf R_{t}})}( - \beta ).
  \label{h}
\end{equation}
Here we separate the sums over hexagon and triangle currents,
$ J({\bf R_{h}})$ and $J({\bf R_{t}})$, with centers ${\bf R_{h}}$ and
${\bf R_{t}}$, and $h'$ numbers three hexagons surrounding each
triangle ${\bf R_{t}}$.

We consider sums of the triple products of the Bessel functions that
appear in Eq.~(\ref{h})
and recall that $I_{n}(- \beta)$ is $n$-th Fourier harmonic of 
$e^{- \beta \cos \phi}$. 
In the low-temperature limit 
($ \beta \gg 1$), 
the integrand in each triple product has extrema at 
${\phi}_{h} = 2 \pi {\sigma}_{h}/3,$ ${\sigma}_{h} = \pm 1$ 
($h = 1,2,3$). Thus, 
one arrives at the following asymptotic formula
\begin{eqnarray}
\sum_{ J({\bf R_{t}})} \prod_{h' = 1}^{3}
 I_{J({\bf R_{t} + A_{h'} }) + J({\bf R_{t}})}( - \beta ) \sim
\nonumber \\
\sum_{\sigma({\bf R_{t}}) = 
\pm 1}\exp\left\{ \frac{
i 2 \pi \sigma({\bf R_t})}{3}
\sum_{h' = 1}^{3} J_{\bf R_{t} + A_{h'} }
\right.
\nonumber \\
\left. 
- \frac{1}{3 \beta}
\sum_{h' > h \geq 1}^{3} 
(J_{\bf R_{t} + A_{h'} } - J_{\bf R_{t} + A_{h} })^{2}
\right\}
\left( 1  + O(\frac{1}{\beta}) \right).
\label{j}
\end{eqnarray}
We note that in addition to hexagon currents, new variables, $\sigma =
\pm 1$, which reside in triangles, appear. These variables count the multiple
ground states.
Now we return to calculation of the partition function (\ref{h}). We
substitute the asymptotic formula for the triple products of Bessel
functions (\ref{j}) into Eq.~(\ref{h}) and rewrite the product of
exponentials over the triangles as an exponential of the sum over the
triangles.  We also use the Poisson summation formula.
Finally, we arrive at the following expression for the
partition function
\begin{eqnarray}
  \label{l}
  Z( \beta ) = \sum_{ \{\sigma{( {\bf R_{t} } )} \}, 
\{ m{( {\bf R_{h} } )} \} }
\int 
\exp \left[ 2 \pi i \sum_{\bf R_{h} } J({\bf R_{h} })
 Q({\bf R_{h}})
\right. \nonumber \\
\left. 
- \frac{2}{3 \beta} \sum_{\bf R_{h}, B_{h}}
\left( J({\bf R_{h}}) -  J({\bf R_{h} + B_{h}})
\right)^{2} \right] \prod_{\bf R_{h}} 
{\rm d} J({\bf R_{h}})
,  \\
  \label{m}
  Q({\bf R_{h}}) = m({\bf R_{h}}) + \frac{1}{3} \sum_{\bf A_{t}}
\sigma{({\bf R_{h} + A_{t}})} 
\end{eqnarray}
Here ${\bf A_{t}}$ runs over all six triangles surrounding each
hexagon with the centers ${\bf R_{h}}$, ${\bf B_{h}}$ are six vectors
that connect the centers of nearest hexagons. 
Note that centers of hexagons form a triangular lattice
which is dual to the hexagonal lattice. 

Now one can integrate the partition function (\ref{l}) over the
currents in hexagons, $J({\bf R_{h}})$. This
results in the expression for the partition function of the 2D Coulomb
gas with quasi-charges  $Q({\bf R_{h}})$ (Eq.~(\ref{m}))
positioned on sites of the triangular lattice ${\bf R_{h}}$. 
Quasi-charges are  1/3-multiple, this corresponds to the 
$2 \pi /3$-multiplicity of vortex rotations.

At zero temperature, the integration over $J({\bf R_{h}})$ in (\ref{l})
yields conservation conditions $\prod_{\bf R_{h}} \delta{(Q({\bf
    R_{h}}))}$, i.~e., in any ground state, the sum of
chiralities of triangles surrounding each hexagon is a multiple of
3. The problem of counting ground
states is  mapped onto that of coloring of the hexagonal lattice \cite{HR}
which was solved exactly \cite{BX}.
The exact number of ground states, $Z_{N}$, is
equal to $ 1.46099^{N/3} $, where $N$ is the number of
spins. A naive approximation that assumes that
chiralities of triangles surrounding each hexagon are independent
and equally probable gives a good estimate
$
Z_{N} \approx {(11/8)}^{N/3} = {1.375}^{N/3}
$
for the number of the ground states. In this estimate we neglect
correlations between chiralities of triangles surrounding neighboring
hexagons and farther correlations. The effect of those correlations
can be estimated as the inverse number  of the nearest neighbors on
the triangular lattice, $1/6$.

At finite temperatures, excitations against ground states,
vortices with nonzero quasi-charges $Q$, appear. Both integration over
hexagon currents, $J({\bf R_{h}})$, and 
summation over chiralities of triangles surrounding each hexagon
contribute to the probability of vortex formation. Therefore, in
addition to the standard estimate of the probability of vortex formation in
unfrustrated XY magnets, one has to find the contribution due to
various chirality configurations. This accounts for the high degeneracy
of ground states in the Kagom\'{e} antiferromagnet. The lowest energy
vortices have quasi-charges $Q = \pm 1/3$. States with the sum of
chiralities of triangles surrounding a certain hexagon equal to
$\pm 1$ and $\pm 4$ contribute to formation of such $Q = \pm 1/3$
vortices. The number of such configurations, $Z_{1,N}$, differs from
the number of ground states, $Z_{N}$ by some numerical factor,
$w_{1}$. We estimate the factor $w_{1}$ the same naive way as we
estimated the number of ground states, i.~e. we assume that 
chiralities $\pm 1$ have equal and independent probabilities, 
$
w_{1} \approx {21/22}.
$
The precision of this estimate is again of order $1/6$.

Thus, the low-temperature partition function is the sum over
states with zero quasi-charge and over states with quasi-charges equal
to $\pm 1/3$. It is convenient to redefine $\Psi{({\bf R_{h}})} =
J{({\bf R_{h}})}/(3 K)$ and $K = {\beta}/12$. In terms of new fields
$\Psi{({\bf R_{h}})}$, the partition function reads
\begin{eqnarray}
Z = \int 
e^{
- {K}/{2} \sum_{\bf R_{h}, B_{h}}
{\left[ \Psi{({\bf R_{h}})} - \Psi{({\bf R_{h} + B_{h}})}
 \right]}^{2} }
\nonumber \\
\times
\left[ 1 + 2 w_{1} \sum_{\bf R_{h}} \cos{ \left( 2 \pi K 
\Psi{({\bf   R_{h}})}
\right) 
} \right]
\prod_{\bf R_{h}} {\rm d} \Psi{({\bf R_{h}})}.
  \label{p}
\end{eqnarray}
This expression coincides with that obtained by JKKN
\cite{JKKN}. The quantity $w_{1}$ plays the role of magnetic field
$y_{0}$ in Ref. \onlinecite{JKKN}. In our case, the lattice 
${\bf R_{h}}$ is dual to the lattice of hexagons,
which corresponds to the initial hexagonal lattice. Hence,
the problem of the KT transition on the Kagom\'{e} lattice with the  
antiferromagnetic interaction is mapped onto that of the KT
transition in a ferromagnet on the hexagonal lattice.

In order to find the transition temperature and to estimate the role
of chiralities we extract the short-distance part of the Green's
function in the hexagonal ferromagnet (\ref{p}). In the
continuous limit,  we obtain 
\begin{equation}
Z = \int {\rm D} \Psi{({\bf r})} 
e^{
 - \int {\rm d}^{2} {\bf r} \left[
\frac{\sqrt{3} K}{2} 
{ \left( \nabla \Psi  \right)}^{2} - 
 h a^{-2} \cos{( 2 \pi K \Psi ) }    \right]} ,
\label{q} 
\end{equation}
where $h = 2 w_{1}  e^{- K \pi^2 /2} =  
2 w_{1} e^{- \beta \pi^2 /24}.$
At the KT transition temperature, this is a small field,
therefore it can be neglected according to reasoning by JKKN. 
From the usual renormalization \cite{CL,N,JKKN} we find that the KT
transition on the hexagonal lattice occurs when  
$ \sqrt3/(2K_c) = \pi/2$, i.~e. 
\begin{equation}
T_{c}/J  S^{2} =  \sqrt{3} \pi /72 =  0.0756.
\label{r}
\end{equation}

Note that recent Monte Carlo simulations of the KT transition in the
Kagom\'{e} antiferromagnet \cite{R} yield $T_{c}/J  S^{2}  \approx 0.078
$. This is in a very good agreement with our exact result (\ref{r}).

The existence of a new set of variables, chiralities, qualitatively changes
the spin correlation function compared to that in ``normal'' XY
magnets. Returning to the initial formulation of the problem (\ref{a}),
we consider the correlation function 
${\cal K}(r_0)= \langle
\exp \left(
i\theta(0)-i\theta({\bf r}_0) \right) \rangle $.
In terms of the 
integer-valued variables,
$n({\bf r},{\bf a})$ (see (\ref{d})), we arrive at an
expression  that differs from (\ref{d}), only by arguments of the
$\delta$-functions. Namely, for sites $0$ and ${\bf r_{0}}$ we get 
\begin{equation}
\sum_{\bf a} n({\bf 0,a})=-\sum_{\bf a} n({\bf r}_0,{\bf a})
 = 1.
  \label{ff}
\end{equation}
instead of the conservation condition  (\ref{f}).
This condition is equivalent to a pattern of currents which is a
superposition of currents $J({\bf R_h})$, that flow in the Kagom\'{e} 
lattice and obey the condition (\ref{f}), and an unit current which 
is created in the point ${\bf 0}$ and is
annihilated in the point ${\bf r}_0$. 
Thus, the correlation function
${\cal K}(r_0)$ has the form 
\begin{eqnarray}
{\cal K}(r_0)=
\frac{1}{Z({\beta})}\sum_{\{ J({\bf R}) \} }
\prod_{\bf (R\neq R^*,A\neq A^*)} \nonumber \\
I_{J({\bf R + A}) + J({\bf R})}( - \beta )
\prod_{\bf (R^*,A^*)}
I_{J({\bf R^* + A^*}) + J({\bf R^*})+1}( - \beta ),
  \label{gg}
\end{eqnarray} 
analogous to that of Eq.~(\ref{g}). Here ${\bf (R^*,A^*)}$ are sites
and vectors of the dual lattice such that ${\bf A^*}$ crosses the path 
$({\bf 0,r}_0)$ on the initial kagom\'{e} lattice.
Now we integrate over currents in the triangles in Eq.~(\ref{gg}) as
we did it before using the asymptotic formula  (\ref{j}). Instead of
Eq.~(\ref{l}) we get
$
{\cal K}(r_0) = {Z( {\beta}, r_0 )}/{Z( \beta )}
$
where ${Z( \beta )}$ is given by Eq.~(\ref{l}) and $Z( {\beta}, r_0 )$
differs from ${Z( \beta )}$ by the additional contribution from the
unit current running along the path $({\bf 0,r}_0)$.

The vertex contribution in the large $r_0$ asymptotics of  the spin
correlation function  ${\cal K}(r_0)$ below the KT-transition point is
negligible because the renormalization-group flow at $T<T_c$ 
yields that the effective constant $h$ in (\ref{q}) 
is equal to zero. This is equivalent
to the neglect of the first term in square brackets in
Eq.~(\ref{l}) in the expression for $Z( {\beta}, r_0 )$. 
Neglecting constraints on chiralities of triangles as we did before we
immediately get 
a factor 
$(\cos 2\pi/3)^{r_0/a}=(-1)^{r_0/a} 2^{-r_0/a}$
in the correlation function,  
where $a$ is the Kagom\'{e} lattice constant. 
The 
integration over $J({\bf R_{h}})$ in the $r_0\to\infty$ limit 
can be done in the spin-wave approximation and the result
coincides with that by JKKN  \cite{JKKN}. Thus,  in the
low-temperature phase $T\leq T_c$ in the long-distance 
limit $r_0/{a} \gg 1$ the spin correlation function reads 
\begin{equation}
{\cal K}(r_0) \propto 
(-1)^{r_0/{a}} 2^{-r_0/{a}}
(r_0/{a})^{-{T}/{36 T_c}} .
 \label{kr}
\end{equation}
It decays exponentially with distance. The true order parameter of
the KT-transition is the cubed  spin \cite{HR}, 
$\eta({\bf r})=\exp(3i\theta({\bf r}))$. 
The correlation function of this order parameter at $T<T_c$
decays as a power of distance
\begin{equation}
<\eta(0)\eta({\bf r}) > \sim
(r_0/{a})^{-{T}/{4T_c}}.
 \label{krr}
\end{equation}

In conclusion, it is shown that the XY antiferromagnet on the
two-dimensional Kagom\'{e} lattice exhibits 
a Kosterlitz-Thouless
transition and the transition temperature, $T_{c}$, is evaluated 
(Eq.~(\ref{r})). It is found that the spin correlation function decays
exponentially with distance even below the transition
temperature (Eq.~(\ref{kr})). Nevertheless, the order parameter
correlations decay as a power of the distance in the low-temperature
phase  (Eq.~(\ref{krr})).

\acknowledgments 

This research 
was supported 
by grants \# NPF000 and \# NPF300 from the International  
Science Foundation and the Russian Government, by the 
US --- Israel Binational Science Foundation,
and by the Russian Foundation for Fundamental Research
(grant \# 96-02-19125a).

\end{document}